# Thermal conductivity of boron arsenide above 2100 watts per meter per Kelvin at room temperature

Ange Benise Niyikiza[a1], Zeyu Xiang[b1], Fanghao Zhang[b1], Fengjiao Pan[a], Chunhua Li[c], David Broido[c], Ying Peng[a*], Bolin Liao[b*], and Zhifeng Ren[a*]

[a]Department of Physics and Texas Center for Superconductivity at the University of Houston (TcSUH), Houston, Texas, USA.

[b]Department of Mechanical Engineering, University of California at Santa Barbara, Santa Barbara, California, USA.

[c]Department of Physics, Boston College, Chestnut Hill, Massachusetts, USA.

[1] The authors contributed equally to this work

*Corresponding authors: Ying Peng (ypeng23@central.uh.edu), Bolin Liao (bliao@ucsb.edu), Zhifeng Ren (zren@uh.edu)

## Abstract

Boron arsenide (BAs) single crystals had been previously reported to have thermal conductivity of 1500 W/m·K at room temperature. Now we achieved thermal conductivity above 2100 W/m·K at room temperature in BAs crystals due to much lower concentration of impurities Si, C, and O grown from purified arsenic. We also observed a $T^{-1.8}$ dependence of the thermal conductivity, suggesting a more significant contribution from four-phonon scatterings than suggested by previous theory. We found that our experimental result can be fit with a modified theoretical calculation by tuning down the three-phonon scattering for phonons in the 4-8 THz range, although current phonon transport theory cannot provide a physical explanation. Such an advance will not only attract more effort on growing BAs single crystals and studying their practical applications but also stimulate theoretical work to predict more materials with possibly even higher thermal conductivities.

# Introduction

Research on boron arsenide (BAs) before 2013 was rare [1–3]. However, ever since the prediction of thermal conductivity of 2200 W/m·K at room temperature in BAs in 2013 assuming only three-phonon scattering plays the dominant role [4], growing BAs single crystals has been explored extensively. BAs crystals with lots of defects were grown and showed thermal conductivity from 200 to 400 W/m·K [5–7]. Later theoretical efforts showed that four-phonon scattering significantly reduced the theoretical BAs thermal conductivity from 2200 to about 1400 W/m·K at room temperature (RT) [8]. Subsequently, these predictions were confirmed by experiments with measured RT thermal conductivities from 1000 to 1300 W/m·K in 2018 [9–11]. Subsequent efforts have focused on the growth of large BAs crystals for exploring a broad range of applications utilizing the high thermal conductivity. However, we noticed that the crystals showing RT thermal conductivity of 1200 W/m·K still had a high concentration of point defects, including Si, C, O, *etc*. [12]. This suggests the possibility that the intrinsic RT thermal conductivity of BAs could be higher than 1200 W/m·K if a much lower point defect concentration could be achieved. Following this line of reasoning, we have worked to decrease or even eliminate the impurities during growth. Experimentally, we found that our previous As source does have a high concentration of Si, C, O, *etc*. Through physical and chemical methods, we have been able to reduce these impurities from the as-received As source material. In our previous effort using the partially purified As, we had grown BAs single crystals with much lower impurities and had experimentally measured thermal conductivity higher than 1500 W/m·K in 2024 [13]. Recently, we have made the source As even purer by further eliminating impurities. With the newly purified As, we have grown BAs crystals with a significantly higher measured thermal conductivity at RT. Using the standard time-domain thermoreflectance (TDTR) method, we have measured RT thermal conductivity around 2200 W/m·K on many spots on multiple crystals from different batches. Since the measured thermal conductivity is much higher than the previously predicted value of 1280 W/m·K [9], we examined possible modifications of the phonon transport theory and we found that by significantly suppressing the three-phonon scattering for phonons in the 4-8 THz range, the calculated thermal conductivity of BAs became close to the measured RT value of 2100 W/m·K, and its temperature dependence from 300 to 400 K follows $1/T^{\alpha}$ with $\alpha$ of about 1.8, which is consistent with our experimentally measured $\alpha$ value of 1.81. This realization of measured RT thermal conductivity above 2100 W/m·K will introduce a new era for BAs research and accelerate the growth effort to obtain larger crystals for a variety of applications. Furthermore, our result will spur theoretical research on phonon transport to understand the higher measured thermal conductivity in BAs presented here and to discover more high-thermal-conductivity materials.

**Crystal Growth**

High-quality boron arsenide (BAs) single crystals were synthesized via a four-step process. First, a quartz tube was cleaned using a standard cleaning procedure for semiconductors, involving sequential ultrasonic cleaning with acetone, ethanol, and deionized water, followed by baking in an oven to eliminate residual moisture, then the arsenic source was purified to minimize impurities such as Si, C, O, *etc*. to ensure high purity of the starting As [14]. Second, purified arsenic (>99.9999%, Alfa Aesar), boron (99.9999%, Alfa Aesar), and iodine (99.99%, Sigma-Aldrich) were loaded into the quartz tube within a glovebox maintained at < 0.1 ppm oxygen and < 0.1 ppm water. The use of high-purity arsenic is critical for reducing the impurities of Si, C, O in the final BAs single crystals so that the reduction of thermal conductivity of BAs by point defects can be minimized [12]. The tube was then sealed under a vacuum of $10^{-4}$ Torr. Third, the sealed quartz tube was placed in a two-zone tube furnace, with the hot zone at 1173 K and the cold zone at 1123

K, for a growth duration of 10-14 days. Finally, BAs single crystals were picked up after breaking the sealed quartz tubes (see Supplementary Information).

**Structure and Crystallinity**

We obtained optical micrographs of representative BAs crystals exhibiting high thermal conductivity, under transmission light (Fig. 1a) and reflection light (Fig. 1b), alongside a crystal with lower thermal conductivity in transmission light (Fig. 1c). These images reveal non-uniformities, evidenced by contrast variations in Fig. 1a and Fig. 1c. It is worth noting that there is a twin boundary and surface irregularities in Fig. 1b. Structural characterization was performed using conventional X-ray diffraction (XRD) (Fig. 1d) and a high-resolution single-crystal diffractometer (Bruker D8 Venture) (Fig. 1e and 1f). The XRD pattern confirms a single-phase zinc-blende cubic face-centered structure oriented along the [111] direction. Single-crystal diffraction patterns of the high thermal conductivity sample (Fig. 1e) display sharp, uniformly spaced spots arranged in a square lattice, free of any other spots, indicating high crystalline quality with negligible Si, C, and O impurities. In contrast, the diffraction pattern (Fig. 1f) of the low thermal conductivity crystal (Fig. 1c) shows there are multiple extra weak spots, suggesting existence of defects that caused the significant reduction on the thermal conductivity.

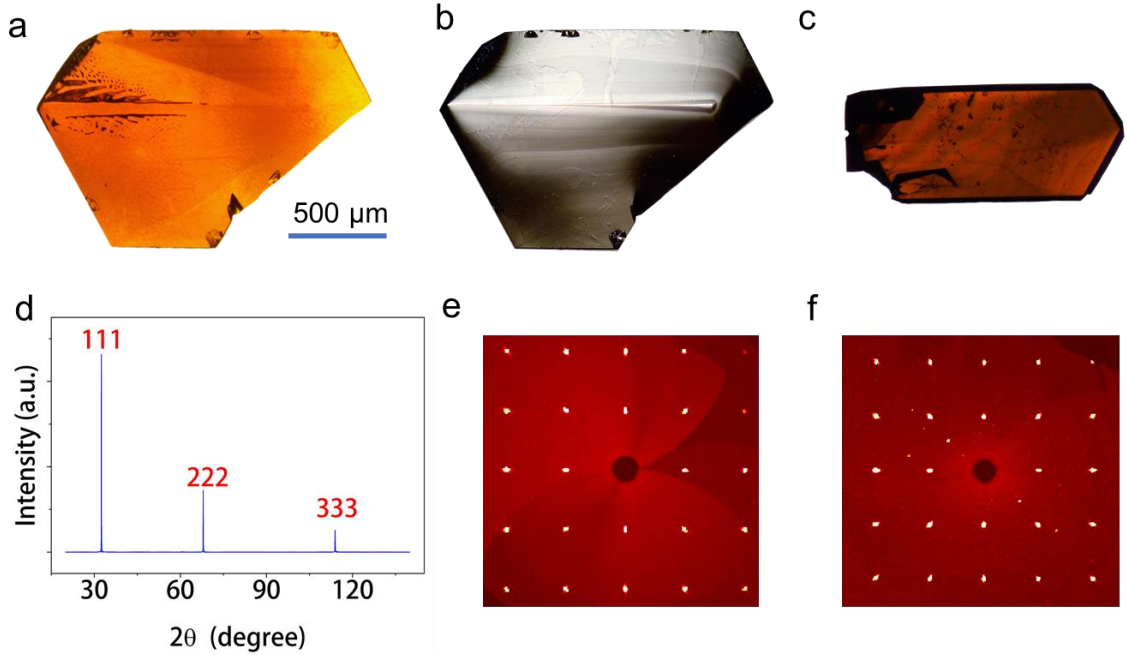

**Fig. 1. Structural characterizations of boron arsenide (BAs). a**, Optical image under transmission light and **b**, under reflection light of a high-quality BAs. **c**, Optical image under transmission light of a low-quality BAs. **d**, 2θ XRD pattern of BAs crystals. **e**, Single crystal x-ray diffraction patterns of a high-quality BAs. **f**, Single crystal x-ray diffraction patterns of a low-quality BAs.

**Thermal conductivity**. We used TDTR to measure the thermal conductivity of our BAs samples. The details about our two-tint TDTR setup are provided in the Supplementary Information. The BAs samples were coated with a 100-nm Al transducer layer using electron beam evaporation before the TDTR measurements. We fit the TDTR phase signal to a heat transfer model to extract the thermal conductivity [15]. Using TDTR, we identified regions in our BAs samples with thermal conductivity as high as 2100 W/m·K at RT. This value is close to that of high purity synthetic single crystalline diamond with natural isotope composition [16]. In Fig. 2a, we show representative TDTR phase signals for a high-thermal-conductivity region in BAs measured with different TDTR modulation frequencies (5 MHz and 10 MHz) and pump beam sizes (diameters of 70 μm, 35 μm, and 19 μm), together with the model fits. The extracted thermal conductivity (~2100 W/m·K) and Al/BAs interfacial thermal conductance (~70 MW/m²·K) are relatively independent of these measurement parameters. The lack of the pump size dependence of the extracted thermal conductivity down to 19 μm indicates that, even in the BAs regions with the highest thermal conductivity, the majority of heat conduction is mediated by phonons with a mean free path below 19 μm, which is consistent with previous calculations [9].

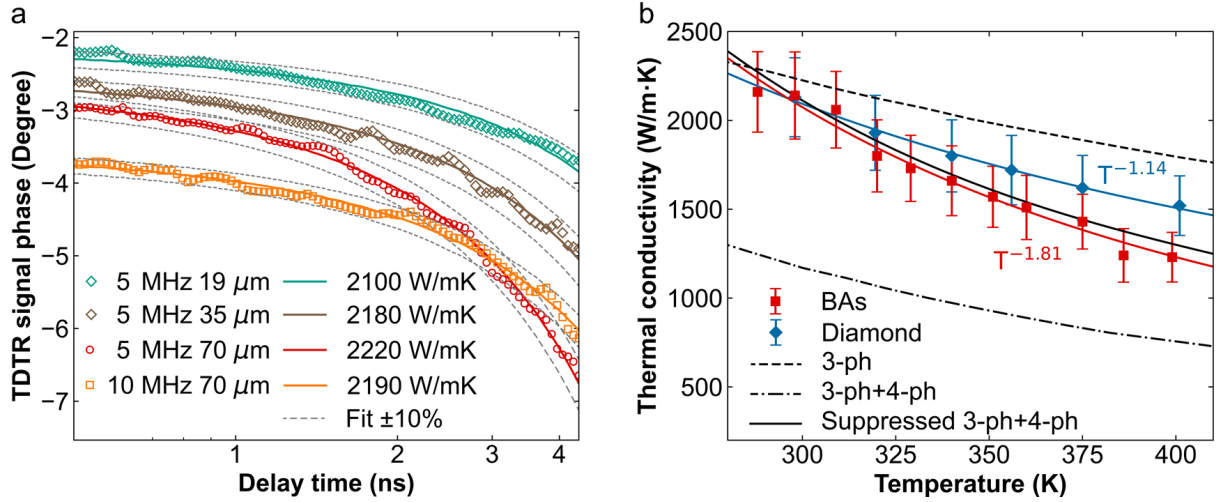

**Fig. 2. Thermal conductivity measurement by time-domain thermoreflectance (TDTR). a**, The TDTR phase signals measured with different pump modulation frequencies (5 MHz and 10 MHz) and pump beam diameters (70 μm, 35 μm, and 19 μm) at high-thermal-conductivity regions of our BAs sample. The solid lines show fitting curves using a heat transport model with the corresponding fitted thermal conductivities labeled in the legend. The dashed lines provide the boundaries of the fitting model when the thermal conductivity is changed by ±10% from the fitted value. **b**, Temperature dependence of the measured thermal conductivity in diamond and BAs single crystals in the temperature range of 290 K to 400 K. The solid blue and red curves represent $1/T^\alpha$ fits with $\alpha$=1.14 for diamond and 1.81 for BAs, respectively. "3-ph": first-principles calculation of the thermal conductivity in BAs with three-phonon scatterings only. "3-ph+4-ph": first-principles calculation of the thermal conductivity in BAs with both three-phonon and four-phonon scatterings. "Suppressed 3-ph+4-ph": a modified calculation with suppressed three-phonon scattering for phonons in the 4-8 THz range.

In Fig. 2b, we show the temperature-dependent thermal conductivity of our BAs sample measured using TDTR. For this measurement, the sample was mounted on a Peltier module and a thermocouple was glued to the surface of the sample to precisely monitor its temperature. We focused on the temperature range of 290 K to 400 K to avoid quasi-ballistic phonon effect at lower temperatures and overestimating the contribution from four-phonon scatterings at higher temperatures. We used the largest pump beam diameter (70 $\mu$m) and both 5 MHz and 10 MHz modulation frequencies for this measurement. The data showed no dependence on the modulation frequency. Our data followed a $T^{-1.81}$ dependence between 290 K and 400 K. Previous measurement of our BAs sample with a RT thermal conductivity of 1500 W/m·K showed a $T^{-2}$ dependence in the temperature range of 300 K to 600 K [13] and a $T^{-1.83}$ dependence between 300 K and 400 K (see Supplementary Information). Earlier measurements of BAs samples with a room-temperature thermal conductivity around 1000 to 1300 W/m·K showed a temperature dependence ranging from $T^{-1.3}$ to $T^{-2}$ [9–11]. In comparison, we also measured the thermal conductivity of a high-quality synthetic single crystalline diamond in the same temperature range, which showed a $T^{-1.14}$ dependence (Fig. 2b). The temperature dependence of thermal conductivity depends on the dominant scattering mechanisms of heat carriers (phonons in this case). In defect-free single crystals, dominant three-phonon scatterings should lead to $T^{-1}$ dependence, as exemplified by our diamond data. In contrast, dominant four-phonon scatterings should lead to $T^{-2}$ dependence. Four-phonon scatterings are known to be important (but not dominant) in BAs due to its large acoustic-optical frequency gap. First-principles calculations incorporating both three-phonon and four-phonon scattering processes suggest a RT thermal conductivity around 1500 W/m·K and a $T^{-1.6}$ dependence in single crystalline BAs with natural isotope composition [8,9].

The higher measured thermal conductivity values and stronger temperature dependence compared with the state-of-the-art first-principles phonon calculations is puzzling. In BAs, the high thermal conductivity is achieved in part through features in the phonon dispersions such as a large frequency gap between the heat-carrying acoustic phonons and optical phonons, and the bunching together of the acoustic phonon dispersions [4]. These features result in unusually weak three-phonon scattering rates for acoustic phonons in the 4 to 8 THz range that are comparable to the higher-order four-phonon scattering rates, according to first principles calculations [4,8]. Since the strength of the three-phonon scattering rates should depend sensitively on how closely the acoustic branches are bunched together in the 4-8 THz [17], as a simple test, we artificially suppressed the three-phonon scattering rates for this group of phonons in our first-principles calculation (details in the Supplementary Information; results plotted in Fig. 2B) and found that both the RT thermal conductivity value and its temperature dependence are in better agreement with the experiment. However, we emphasize here that this modified calculation is only meant to show that reduced three-phonon scatterings can potentially lead to better agreement with the experimental result. Given the lack of tunable parameters in state-of-the-art first-principles phonon transport calculations, our experimental result cannot be explained by existing theory and calls for new theoretical work to help understand the higher measured BAs thermal conductivity presented here.

**Conclusion**

We achieved an unprecedented measured thermal conductivity of 2100 W/m·K in boron arsenide (BAs) single crystals at RT, significantly surpassing prior theoretical and experimental benchmarks. By employing ultrapure arsenic and an optimized four-step synthesis, we drastically reduced Si, C, and O impurities to further raise the limit of BAs's intrinsic thermal conductivity, challenging existing theoretical predictions. This record-breaking result, rivaling that achieved in high-purity synthetic diamonds, underscores our breakthrough in crystal growth and purification. It positions BAs as a transformative material for the semiconductor industry, promising superior thermal management in high-power electronics and optoelectronics. Our findings spur further efforts in large-scale crystal growth and theoretical exploration of ultrahigh-thermal-conductivity materials, paving the way for next-generation technologies.

**Acknowledgments:**

**Funding:** This work is supported by the U.S. National Science Foundation Future of Semiconductors (FuSe) Program under the award number DMR-2425439. C.L. and D.B. were supported by the US Department of Energy, Office of Science, Basic Energy Sciences, under award no. DE-SC0021071 (*ab initio* calculations of the phonon thermal conductivities).

**Author contributions:** Z.R. and B.L. lead the project; A.N., Y.P., F.P. synthesized the materials and performed single-crystal analysis; B.L., Z. X., and F.Z. performed TDTR measurements; C.L. and D.B. calculated the BAs thermal conductivity with suppressed three-phonon scattering. All authors edited the manuscript.

**Competing interests:**

Authors declare that they have no competing interests.

# Supplementary Materials

**Thermal conductivity of boron arsenide above 2100 watts per meter per Kelvin at room temperature**

Ange Benise Niyikiza[a1], Zeyu Xiang[b1], Fanghao Zhang[b1], Fengjiao Pan[a], Chunhua Li[c], David Broido[c], Ying Peng[a*], Bolin Liao[b*], and Zhifeng Ren[a*]

[a]Department of Physics and Texas Center for Superconductivity at the University of Houston (TcSUH), Houston, Texas, USA.

[b]Department of Mechanical Engineering, University of California at Santa Barbara, Santa Barbara, California, USA.

[c]Department of Physics, Boston College, Chestnut Hill, Massachusetts, USA.

[1] The authors contributed equally to this work

*Corresponding authors: Ying Peng (ypeng23@central.uh.edu), Bolin Liao (bliao@ucsb.edu), Zhifeng Ren (zren@uh.edu)

**Materials and Methods**

**Crystal growth process**

**(1) Preparation of quartz tubes.** Quartz tubes are used for the growth. Therefore, it is highly important to clean them as they contain residues that could get into the final BAs crystals to reduce the thermal conductivity. Quartz tubes are first observed under the optical microscope to avoid any cracks that can break during the growth, then they are fused and cleaned to remove both organic and inorganic residues. After cleaning, they are put in the oven to dry.

**(2) Arsenic purification.** Arsenic is purified to reduce impurities that exist in the source material. Even though the arsenic source material has high purity, it still has impurities such as Si, C, and O that can be removed by heating and vapor transfer (see Fig. S1). After completing this step, we use the purified arsenic as As source to grow BAs.

**(3) Ingredients loading.** All tools and ingredients (purified arsenic, iodine, and boron) were transferred into the glove box with oxygen level of < 0.1 ppm and water level of < 0.1 ppm to prevent any air contamination with the air-sensitive material like arsenic. Boron (99.9999%, Alfa Aesar), iodine (99.99%, Sigma Adrich), and purified arsenic (> 99.9999%, Alfa Aesar) were loaded into a 10 cm fused quartz tube and sealed with the quick connect flange that allows them to be transferred out without any further contamination. Under the vacuum of $10^{-4}$ Torr, the quartz tube was fused under the hydrogen torch.

**(4) Crystal growth.** The sealed quartz tube containing all the ingredients was put in a two-zone furnace with hot side at 1173 K and cold side at 1123 K for 10-14 days. After growth, the quartz tube was cooled to room temperature, and the tube was broken to pick up the crystals.

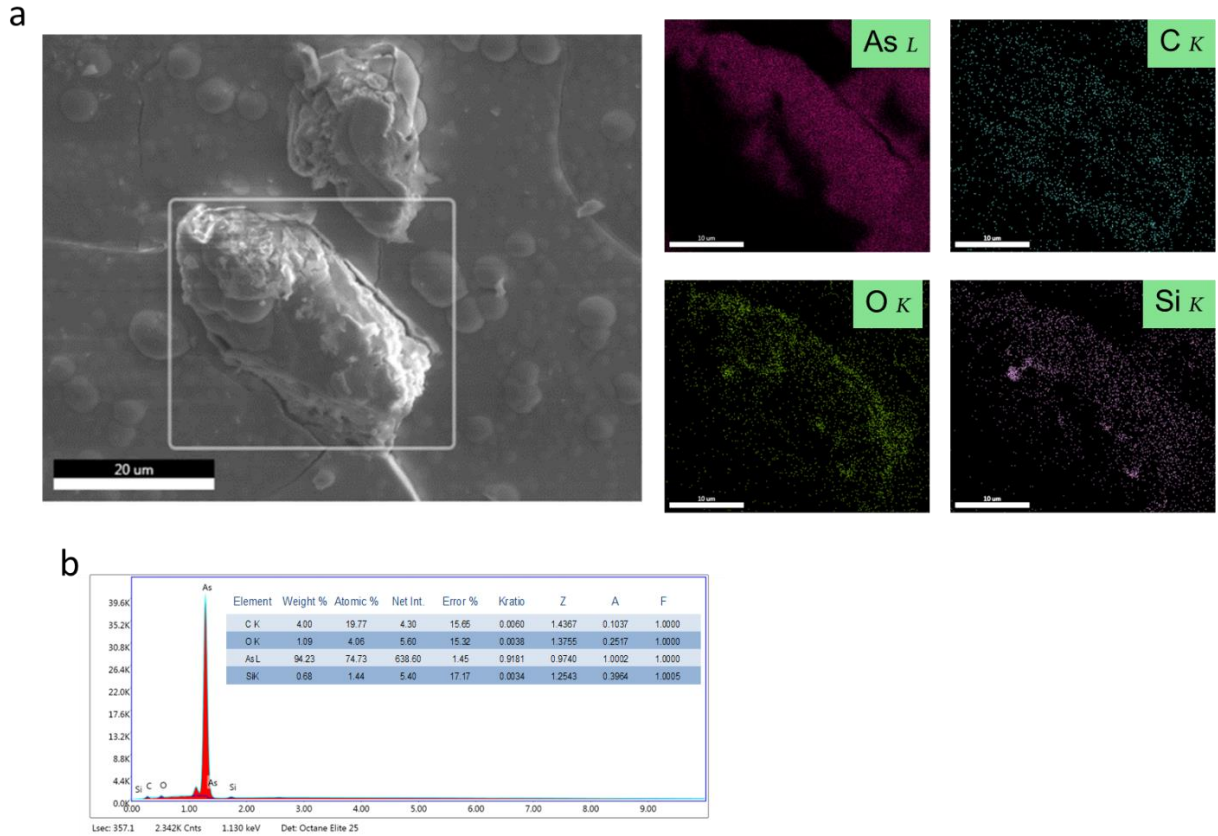

**Fig. S1. Impurities found in the As source. a,** SEM and EDS images of the impurities. **b,** The relative content of impurities.

**Time-domain thermoreflectance**

The thermal conductivity of boron arsenide (BAs) was characterized using time-domain thermoreflectance (TDTR) [1,2]. A 100-nanometer-thick aluminum transducer layer was deposited onto the sample surface via electron beam evaporation, and the thickness of the aluminum film was independently validated through picosecond photoacoustic measurements. The coated transducer layer was first heated up by a pump laser pulse, while a probe pulse subsequently probes the cooling process as a function of the delay time. The entire sample was scanned using TDTR measurements to identify the location exhibiting the highest thermal conductivity, determined by fitting the phase signal based on the in-phase ($V_{in}$) and out-of-phase ($V_{out}$) signals as a function of time delay. A multilayer Fourier heat transfer model is employed to extract both the thermal conductivity and the interfacial thermal conductance between the transducer layer and the sample.

The experimental setup is illustrated schematically in Fig. S2. A mode-locked Ti:sapphire oscillator is employed as the excitation source, producing femtosecond laser pulses with a central wavelength of 800 nm at a repetition rate of 80 MHz. The laser pulse is divided into pump and probe beams using a polarizing beam splitter, followed by spectral filtering with sharp-edged optical filters centered at 792 nm for pump and 806 nm for probe, respectively. The pump beam is modulated at 5 MHz or 10 MHz via an electro-optic modulator (EOM), inducing periodic heating of the sample surface to facilitate thermoreflectance measurements. The intensity of the reflected probe beam is detected by a photodetector connected to a radio-frequency lock-in amplifier, which captures the reflectance variations of the aluminum transducer. A mechanical delay stage precisely controls the relative optical path length difference between the pump and probe beams, allowing for the acquisition of thermoreflectance signals over a temporal delay window spanning from 0 to 5 ns. Using 5 ×, 10 ×, and 20 × objective lenses, the pump and probe beams are focused to $1/e^2$ diameters of 70.0 $\mu$m and 20.0 $\mu$m, 35 $\mu$m and 9 $\mu$m, 19 $\mu$m and 5 $\mu$m, respectively. During the measurements, the pump and probe beam powers are maintained at approximately 20 mW on the sample.

For the temperature-dependent measurement, a commercial Peltier module is employed to heat or cool the sample. The BAs sample is attached to the ceramic surface of the hot side Peltier module using thermal grease, while the cold side is connected to a heat sink at room temperature using thermal paste. The temperature is monitored using a temperature monitor (DP9800 Omega) with a K-type thermocouple (Omega Engineering) that is directly mounted on the sample surface using GE varnish. The heat capacity data of BAs used in the model fitting across various temperatures were adopted from a previous study [3]. For comparison, the thermal conductivity of a high-quality synthetic single-crystalline diamond was measured over the same temperature range using a similar setup with a 5 MHz-modulated pump beam. All measurements were conducted ten minutes after reaching the target temperature to ensure the thermocouple reading stabilizes and the sample reaches thermal equilibrium.

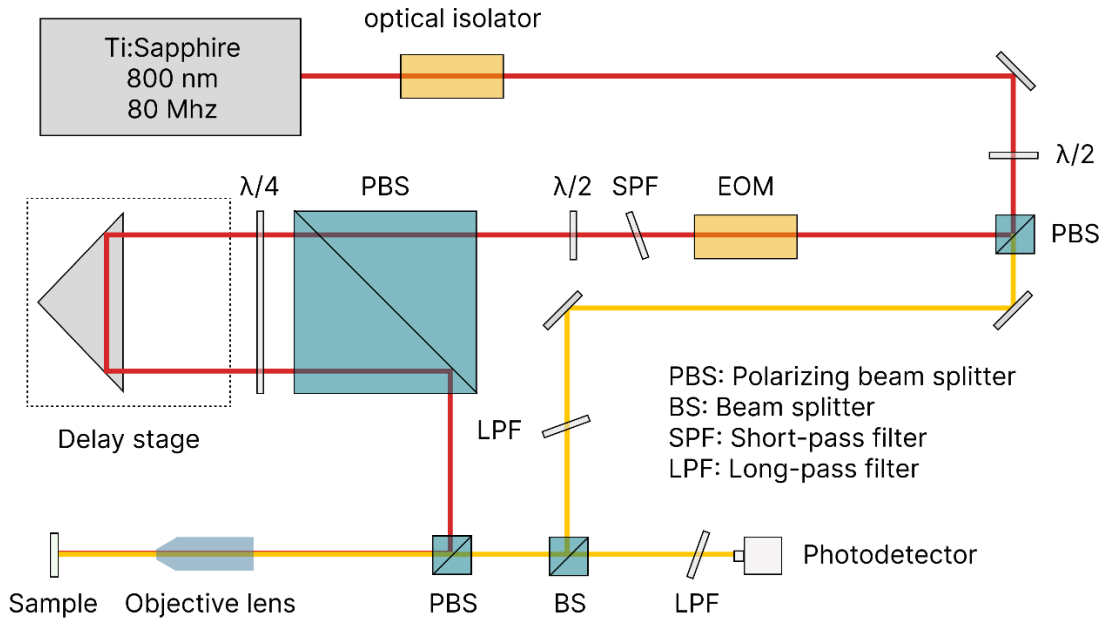

**Fig. S2. The schematic of the TDTR experimental setup.**

To validate the reliability of our TDTR measurements, a sensitivity analysis of both the interfacial thermal conductance $G$ and the thermal conductivity $K$ under varying modulation frequencies and beam sizes is presented in Fig. S3. With an interfacial thermal conductance $G$ around 75 MW/m²·K, using a lower modulation frequency (5 MHz) enhances the accuracy of the thermal conductivity $K$ measurements by increasing the thermal penetration depth, compared to a higher modulation frequency (10 MHz). Although a smaller beam size enhances the phase signal sensitivity to both the thermal conductivity $K$ and interfacial thermal conductance $G$, it also leads to greater uncertainty in beam size determination [4], ultimately increasing the overall measurement uncertainty. The uncertainty of the fitted thermal conductivity $K$ includes typical uncertainties in the control parameters [5], including 3% for transducer heat capacity $C_m$, 5% for transducer film thickness $h_m$ and substrate heat capacity $C$, 10% for beam radius $w$, and 20% for transducer thermal conductivity $K_m$ [5], as calculated using the error propagation formula:

$$\delta K = [(\frac{\partial K}{\partial C_m}\delta C_m)^2 + (\frac{\partial K}{\partial h_m}\delta h_m)^2 + (\frac{\partial K}{\partial C}\delta C)^2 + (\frac{\partial K}{\partial w}\delta w)^2 + (\frac{\partial K}{\partial K_m}\delta K_m)^2]^{\frac{1}{2}}$$

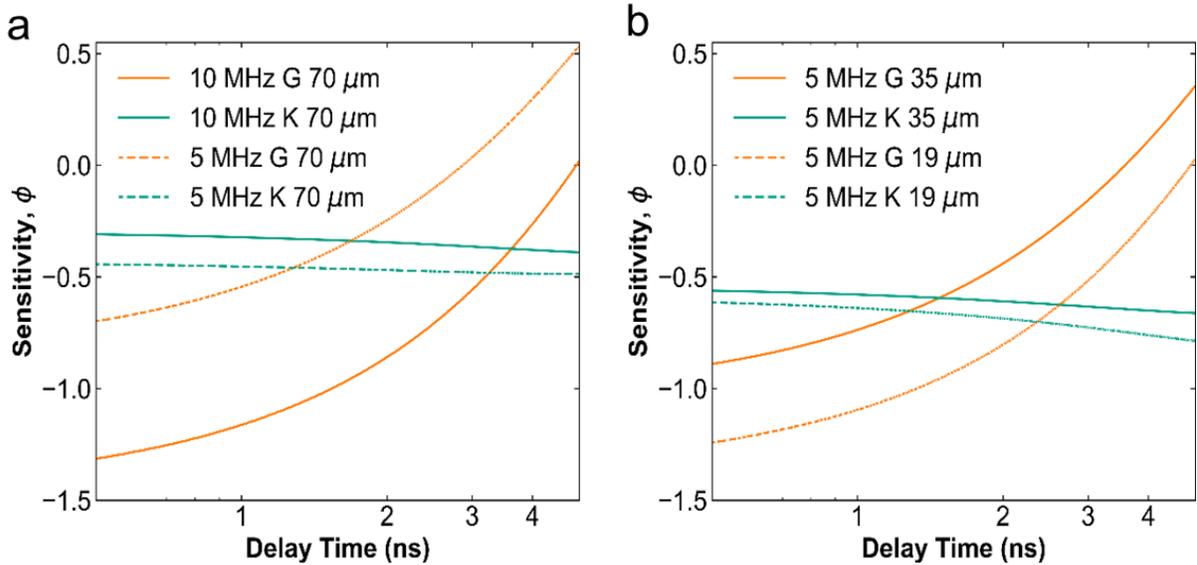

**Fig. S3. Sensitivity analysis for TDTR phase signal of BAs.** The interfacial thermal conductance $G$ and thermal conductivity $K$ are assumed to be 75 MW/m²·K and 2000 W/m·K, respectively, for this analysis. **a**, Under a 5× objective lens (pump beam diameter of 70 μm), the phase signal $\phi$ exhibits greater sensitivity to the thermal conductivity $K$ at a modulation frequency of 5 MHz compared to 10 MHz, whereas the sensitivity to interfacial

thermal conductance $G$ shows the opposite trend. **b,** At a modulation frequency of 5 MHz, the phase signal $\phi$ exhibits increased sensitivity to both the thermal conductivity $K$ and the interfacial thermal conductance $G$ when using a smaller beam size (pump beam diameter of 19 $\mu$m under a 20× objective lens) compared to a larger one (35 $\mu$m under a 10× objective lens).

**First principles thermal conductivity calculation**

Lattice thermal conductivity was computed by iteratively solving the linearized Boltzmann transport equation on a 50×50×50 q grid [6]. We included phonon-isotope, three-phonon and four-phonon scattering. The harmonic interatomic force constants (IFCs) were computed on a 6×6×6 q grid with the density functional perturbation theory [7] implemented in the Quantum-espress suite [8] while third order and fourth order IFCs were obtained using the finite difference method on a 5×5×5 supercell. From these IFCs we computed the phonon frequency, three-phonon and four-phonon scattering rates. To show the effect of reducing three-phonon scattering rates on thermal conductivity we artificially reduced the three-phonon scattering rates in the 4-8 THz range to 1% of the values from first principles.

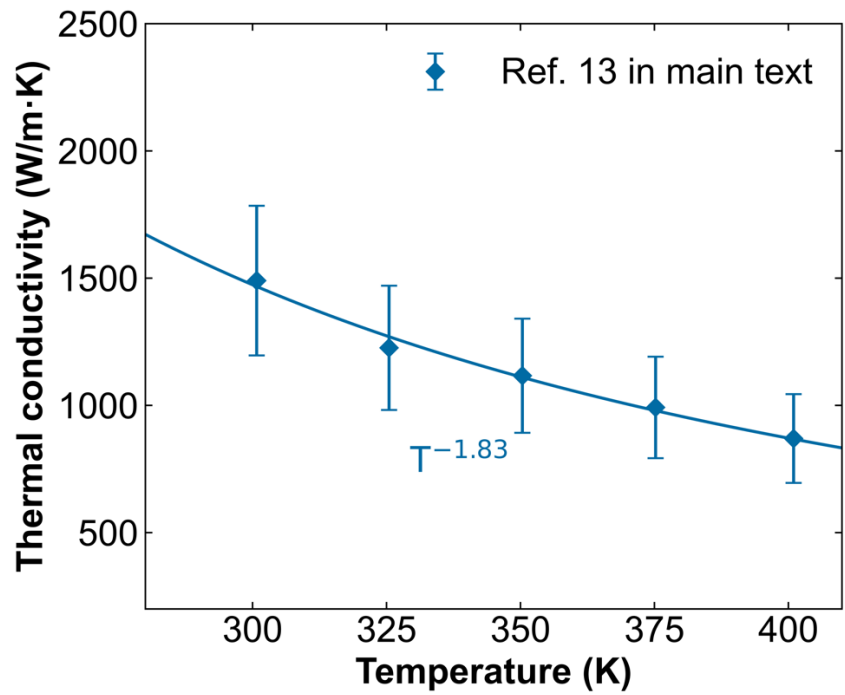

**Fig. S4. Temperature dependence of the thermal conductivity of a BAs sample with RT thermal conductivity of 1500 W/mK between 300 K and 400 K.** Data is taken from Reference 13 cited in the main text.